\documentclass[12pt]{article}

\usepackage{latexsym,amssymb,euscript}
\usepackage[dvips]{graphicx}
\usepackage{amsmath}

\usepackage[symbol]{footmisc}
\usepackage{lineno}
\usepackage{color}
\usepackage{enumitem}
\usepackage{cite}

\begin{document}


\begin{center}
  {\Large \textbf{Finite density $2d$ $O(3)$ sigma model: \break dualization and numerical simulations}}

\vspace*{0.6cm}
\textbf{B. All\'es\footnote{email: alles@pi.infn.it}} \\
\vspace*{0.1cm}
\centerline{\it INFN Sezione di Pisa, Largo Pontecorvo 3, 56127 Pisa, Italy}
\vspace*{0.3cm}
\textbf{O.~Borisenko\footnote{email: oleg@bitp.kiev.ua}} \\
\vspace*{0.1cm}
\centerline{\it N.N.Bogolyubov Institute for Theoretical Physics,} 
\centerline{\it National Academy of Sciences of Ukraine, 03143 Kiev, Ukraine}
\vspace*{0.3cm}
\textbf{A. Papa\footnote{email: papa@fis.unical.it}} \\
\vspace*{0.1cm}
\centerline{\it Dipartimento di Fisica, Universit\`a della Calabria and}
\centerline{\it INFN Gruppo Collegato di Cosenza, Arcavacata di Rende, 87036 Cosenza, Italy}
\end{center}

\begin{abstract}
The action of the $2d$ $O(3)$ non-linear sigma model {on the lattice} in a bath of particles, when expressed in terms of standard $O(3)$ degrees of freedom, is complex.
A reformulation of the model in terms of new variables that makes the action real is presented. This reshaping
enables us to utilize Monte Carlo simulations based on usual importance sampling. 
Several observables, including the {correlation function and the mass gap}, are measured.
\end{abstract}

\section{Introduction} 
\label{intro}

The physics of dense matter is crucial for
understanding the drastic changes underwent by the Universe during the first minutes after the Big Bang. The extreme conditions that prevailed
in that period can be partially reproduced in experiments with heavy-ion collisions where the effects of dense matter are probed.

The presence of a density of matter affects the dynamical laws that govern the physical systems.
The study of these effects by numerical simulations is usually hindered by the so-called sign problem. This problem consists in the fact that upon adding a coupling with an external chemical potential, the
functional that weighs field configurations may not furnish a positive number, thus ruining any attempt to use that functional as a probability for
importance sampling in a Monte Carlo procedure.

Several methods have been proposed to overcome those difficulties in Monte Carlo simulations of Quantum Chromodynamics (QCD), see for instance Ref.~\cite{deforcrand}. The
quest for better strategies has raised the interest in $2d$ toy models that are afflicted by similar problems~\cite{ayyar,bloch,aarts,tanizaki,fujii,mukherjee,cristoforetti,alexandru}.
A technique that has proven to be particularly appealing to evade the sign problem is dualization: the model is recast in terms of
new (dual) variables in such a way that the new action turns out to be real.

No general recipe exists for transforming ordinary into dual variables. Every model has to be studied on its own and the strategy to get a dual representation
may differ significantly for different models. Moreover, not even a unique dual representation exists for a given model. In fact, among the few different possible
dual representations available for a given model, some may be more advantageous than others for simulation purposes or some might even present
a so heavy slowing down that it makes the dual version, albeit real, useless. 

Another type of drawback that often appears in dual formulations is that certain observables cannot be disentangled from the probability 
weight in such a way that their expectation values have to be extracted
as the ratio of expectation values with different Hamiltonians, which
in general gives rise to extremely inefficient numerical calculations. Such difficulties
typically arise in the evaluation of non-local observables like correlation functions.

In the present paper we apply the dualization idea to the $2d$ $O(3)$ non-linear sigma model in presence of a chemical potential.
The standard action of the model on a $d$-dimensional lattice $\Lambda\in \mathbb{Z}^d$ without a chemical potential is given by 
\begin{equation}
  S \ = \ \sum_{x,\nu}\sum_{k=1}^3\sigma_k(x)\sigma_k(x+e_\nu) 
  \label{standardaction}
\end{equation}
{ together with the condition $\sum_k\sigma_k^2(x)=1$ for {every} $x$. The corresponding} partition function reads 
\begin{eqnarray}
Z_{\Lambda}(\beta)  \ = \ \int \prod_{x \in \Lambda} \prod_{k=1}^3 d\sigma_k(x) \  \prod_{x \in \Lambda}
\delta \left ( 1-\sum_{k=1}^3\sigma_k^2(x) \right )  \exp \left [ \beta S \right ]  \; . 
\label{PF_def}
\end{eqnarray} 
In $2d$ this model possesses a particle triplet with a spontaneously generated mass gap $m$~\cite{hasenfratz}. 
The value of this mass has been verified numerically~\cite{kim,alles}. The model is asymptotically free and presents
a rich topology~\cite{polyakov}. All those properties make this model a close relative of QCD. 

The physical interest of the $2d$ $O(3)$ non-linear sigma model goes well beyond the above list of properties. 
In the first place the non-linear sigma model reproduces reliably {several qualitative traits} of ferromagnetic materials. 
Another reason for casting relevance to the model is that it is involved in the development of the resurgence program~\cite{fateev, dunne}.

Our purpose is to construct a real action for the $2d$ $O(3)$ non-linear sigma model at finite density expressed in terms of dual variables and in such
a way that it allows us to determine vacuum expectation values with usual Monte Carlo methods.

To describe the theory at finite density one introduces a chemical potential $\mu$ in the original action 
as an external source for the total third component of the angular momentum~\cite{hasenfratz}. 
Following~\cite{bruckmann} and using the spherical parameterization 
\begin{equation} 
\sigma_1 = \sin\alpha \cos\phi \;,\qquad \ \sigma_2 = \sin\alpha \sin\phi \;,\qquad \ 
\sigma_3 = \cos\alpha \ ,
\label{spher_coord}
\end{equation}
{ we derive} the { standard} action { with a non-zero chemical potential}
\begin{eqnarray}
  S(\{\alpha(x), \phi(x)\})\hspace{-3mm}&=\hspace{-3mm}&\sum_{x,\nu}  \biggl[\cos\alpha(x) \cos\alpha(x+e_{\nu}) \biggr. \nonumber\\
\hspace{-3mm}&+\hspace{-3mm}& \biggl.\sin\alpha(x) \sin\alpha(x+e_{\nu})  \cos\big(\phi(x) - \phi(x+e_{\nu}) - i \mu_{\nu}\big) \biggr] \;,
\label{Sk_def_mu}
\end{eqnarray}
where the general situation with anisotropic chemical potentials $\mu_\nu$, $\nu=1,2$ is considered. 
The conventional physical situation is recovered if we put $\mu_1=\mu,\mu_2=0$.

Three different routes to introduce dual variables are possible, and it is not obvious {\it a priori} 
which one is preferable and under what circumstances may
it be so. The first route relies on the fact that the chemical potential is introduced 
in the Abelian, {\it i.e.} $O(2)$, subgroup. The $O(2)$ part of the general $O(N)$ action has the form 
\begin{equation} 
S = \sum_{x,\nu} \beta_{\nu}(x)  \cos(\phi(x) - \phi(x+e_{\nu}) - i \mu_{\nu}) \ . 
\label{O2_def_mu}
\end{equation} 
This action is of an $XY$ type with a fluctuating coupling and in the presence of $\mu_{\nu}$. 
As shown in~\cite{ogilvie} in the case of the Villain formulation of the $XY$ model and for $\beta_{\nu}(x)=\beta$ for all $x$, 
the conventional dual transformations performed with a non-vanishing chemical potential lead to a positive definite 
Boltzmann weight. Moreover, if the coupling $\beta_{\nu}(x)$ is positive for all $x$, then it is straightforward 
to prove that exactly the same transformations lead to a positive dual weight for all $O(N)$ models. This conclusion has been 
explicitly demonstrated in~\cite{finite_mu_u1o3} for the $2d$ $O(3)$ model, and the proof can be readily 
extended to all $O(N)$ models.

The second route consists in Taylor expanding the Boltzmann weight and integrating
over the original degrees of freedom. The dual variables appear as flux variables subject to certain constraints, and the dual 
weight can be proven to be positive for $O(N)$ models~\cite{finite_mu1,finite_mu2}. The resulting dual theory can be simulated 
by a worm algorithm, and a number of thermodynamic quantities have been computed along this route~\cite{finite_mu1,finite_mu2,niedermayer}. 

In the {third route} one constructs the dual theory by expanding the Boltzmann weight in hyperspherical harmonics on $O(N)$ and 
integrating out the original variables. This program has been accomplished for the $2d$ $O(3)$ model in~\cite{finite_mu_u1o3}. 
However, the full positivity of the resulting dual weight remains to be proven.
It is important to
stress that, at least in the context of these two-dimensional models, the dual formulation appears
as the only reliable tool to investigate the properties of the model. Indeed, the results of Ref.~\cite{niedermayer} show 
that alternative approaches to the sign problem ---the reweighting and the complex Langevin--- have certain drawbacks 
and lead to incorrect results in some regions of the $\beta$-$\mu$ plane. 

One of the main purposes of the present work is to develop a dual formulation applicable not only for computing thermodynamic 
quantities, but also for extracting long-distance quantities like the correlation functions. With such a formulation in hand, one could be 
able to reliably address the question of the hypothetical Berezinskii-Kosterlitz-Thouless (BKT) transition in $O(N)$ models at non-vanishing chemical potential.
Our approach is essentially the first route described above. We reformulate the $O(N)$ model in terms of the link formulation for the Abelian 
(sub)group. In this formulation our results can be straightforwardly extended to all $O(N)$ models in any dimension. 
Here, for the sake of simplicity, we consider only the two-dimensional $O(3)$ sigma model.  

The paper is organized as follows. In the next Section we construct the dual formulation. Firstly we introduce 
the link representation, then we obtain two alternative dual Boltzmann weights, both of which are positive. To study the real effectiveness of the above
Boltzmann weights in Monte Carlo simulations, we test one of them by calculating numerically several observables. We are particularly careful to
individuate any slowing down during the simulations. In Section~\ref{simdetails} we outline the procedure employed during the simulations and list the observables that have been
studied. Specifically we calculate the particle density to find the expected threshold at $\mu=m$, verify that the mass gap extracted from a correlation function is
insensitive to $\mu$, and evaluate the energy as a function of $\beta$ and $\mu$. The results are presented in Section~\ref{results} and some conclusive remarks in Section~\ref{discus}.

\section{Dual representation}  
\label{dualrepre}

{The} action~(\ref{Sk_def_mu}) becomes complex if any of the $\mu_{\nu}$'s is non-zero. 
However, an action with a positive Boltzmann weight can be constructed by using,
in the spirit of~\cite{finite_mu1,finite_mu2}, a dual representation of the model with the action~(\ref{Sk_def_mu}). 
Nevertheless, our strategy is somewhat different from~\cite{finite_mu1,finite_mu2} and relies on the use of the so-called link formulation~\cite{linkrepr,linkform}. In fact, this approach is similar to the one used in~\cite{ogilvie,finite_mu_u1o3}. 
One of its advantages is that it can be readily extended to any $O(N)$ model, in any number of space dimensions.

\subsection{Link formulation for Abelian subgroup}

To build a dual representation, we start from the following partition function with the action~(\ref{Sk_def_mu}),
\begin{eqnarray}
Z_{\Lambda}(\beta,\mu_{\nu})  \ = \  \int_0^{\pi} \prod_x \frac{d \alpha(x)}{2} \sin\alpha(x)\ 
\int_0^{2\pi} \ \prod_x \ \frac{d\phi(x)}{2\pi} \ \exp \left [ \beta S(\{\alpha(x),\phi(x)\} ) \right ]  \; . 
\label{PF_def_mu}
\end{eqnarray} 
The integration can be done in a number of ways. Here we use the fact that the dependence of~(\ref{Sk_def_mu})
on the angles $\phi(x)$ is only through their differences ($U(1)$ variables). This allows to make a change 
of variables and rewrite the partition function in terms of the link angles 
$\phi(l)\equiv \phi_{\nu}(x)= | \phi(x)-\phi(x+e_{\nu}) |_{\mbox{mod} (2\pi)}$,
where links are defined as $l\equiv(x;\nu)=(x_1,x_2;\nu)$ if $d=2$.

The procedure generates local and global constraints known as Bianchi identities on the link variables~\cite{linkrepr,linkform}.  
The local identity constrains the allowed configurations of $\phi(l)$ on every plaquette $p$ of the lattice and can be embedded 
into the partition function in the form of a periodic {$\delta$-function} as 
\begin{eqnarray}
\label{bianchi_local}
\prod_p \ \sum_{r(p)=-\infty}^{\infty} \ e^{i r(p) \phi(p)} \ , \
\phi(p) = \phi(l_1) + \phi(l_2) - \phi(l_3) -\phi(l_4) \ , \qquad l_i\in p \ . 
\end{eqnarray}
Global identities constrain two holonomies winding through the lattice in periodic directions. They have the form
\begin{eqnarray}
\label{bianchi_global}
\sum_{q_1=-\infty}^{\infty} \ \sum_{q_2=-\infty}^{\infty} \ e^{iq_1\sum_{x_1}\phi_1(x_1,0) + iq_2\sum_{x_2}\phi_2(0,x_2)} \ . 
\end{eqnarray} 
Then, it is easy to prove the following equality in any number of dimensions $d$:
\begin{eqnarray}
\label{bianchi}
&&\int_0^{2\pi} \ \prod_x \ \frac{d\phi(x)}{2\pi} \ e^{S(\{\phi(x)-\phi(x+e_{\nu}) - i\mu_{\nu} \} ) } \ = \ 
\int_0^{2\pi} \ \prod_l \ \frac{d\phi(l)}{2\pi} \ e^{S(\{\phi(l) - i\mu_{\nu} \})} \\ 
&&\times \prod_p \ \sum_{r(p)=-\infty}^{\infty} \ e^{i r(p) \phi(p)} \ \prod_{\nu=1}^d \ 
\sum_{q_\nu=-\infty}^{+\infty} \ e^{iq_\nu\sum_{x_\nu}\phi_\nu(0,\dots, x_\nu, \dots, 0)} \ . 
\nonumber 
\end{eqnarray}
One should keep in mind that when $d>2$ not all local Bianchi identities are independent.
The chief argument in favor of using the link formulation lies in the following fact. The dependence of the partition
function on the chemical potential in the finite temperature theory can appear only through loops winding through
the whole lattice in the compactified direction, {\it i.e.}, through Polyakov loops in terms of gauge theories.
In models with a global symmetry group such loops are represented by holonomies which enter in the global Bianchi identities.
Therefore, a dependence on the chemical potential can only appear due to non-zero contributions from global variables
$q_{\nu}$ representing constraints on such holonomies. This is what the formula (\ref{bianchi}) demonstrates. Indeed, making
a global shift of link variables $\phi(l)\equiv\phi_{\nu}(x)\to \phi(l)+i\mu_{\nu}$ and using the periodicity of the integrand
in (\ref{bianchi}), one sees that the chemical potential decouples from the integrand and appears in the partition function
only through global variables $q_{\nu}$ as $e^{-q_{\nu} \mu_{\nu} L_\nu}$. Moreover, this simple transformation brings the partition
function to the form in which the contribution of the chemical potential is always real and positive.

Applying this approach to the $2d$ $O(3)$ model one gets, after integration over link variables, 
\begin{eqnarray} 
Z_{\Lambda}(\beta,\mu_{\nu})  \ = \ \sum_{q_1=-\infty}^{\infty} \ \sum_{q_2=-\infty}^{\infty} \ 
e^{-\sum_{\nu=1,2} \ q_{\nu} \mu_{\nu} L_\nu } \  
\sum_{\{r(p) \}=-\infty}^{\infty} \int_0^{\pi} \prod_x \frac{d \alpha(x)}{2} \sin\alpha(x) \nonumber  \\ 
\times\exp \left [ \beta \sum_{x,\nu}\cos\alpha(x)\cos\alpha(x+e_{\nu}) \right ] 
\prod_{x,\nu} \ I_{r(l)}\left ( \beta \sin\alpha(x)\sin\alpha(x+e_{\nu}) \right ) \ , 
\label{Zsigma_mu}
\end{eqnarray}
where $L_\nu$ are the values of the lattice size in the two directions, $I_{r(l)}$ is the modified Bessel function of first kind, and
\begin{eqnarray} 
r(l) \ = \ 
\begin{cases}
r(p_1)-r(p_2) + q_{\nu} \ , {\rm if} \ l=(x_1,0;1) \ {\rm or} \ l=(0,x_2;2) \ ,    \\  
r(p_1)-r(p_2) \ , \ {\rm otherwise} \ . 
\end{cases}
\label{rl}
\end{eqnarray} 
The plaquettes $p_1$ and $p_2$ have the link $l=(x;\nu)$ in common. 

The two-point correlation function between the origin (denoted by a nought $0$) and a point ${R}$ in the parameterization~(\ref{spher_coord}) reads 
\begin{equation}
  \Gamma({R}) =  \Gamma_1({R}) + \Gamma_2({R}) \;,
  \label{Gamma_zeroteta}
\end{equation}
with
\begin{eqnarray}
\Gamma_1({R})  \hspace{-3mm}&\equiv\hspace{-3mm}&   \langle  \cos\alpha(0) \cos\alpha({R})  \rangle    ,\nonumber         \\  
\Gamma_2({R}) \hspace{-3mm} &\equiv\hspace{-3mm}&   \langle  \sin\alpha(0) \sin\alpha({R})  \cos(\phi(0) - \phi({R}))  \rangle \;,
\label{Gamma_zeroteta1}
\end{eqnarray}
where the expectation values $\langle\cdots\rangle$ are evaluated with~(\ref{PF_def_mu}). In the link formulation $\Gamma_2({R})$ is given by 
\begin{equation}
\Gamma_2({R}) \ = \  \left\langle \  \sin\alpha(0) \sin\alpha({R}) \ \cos\left (\sum_{l\in C_{R}}\eta(l) \phi(l) \right ) \  \right\rangle   \ , 
\label{corr_link}
\end{equation}
where $\eta(l)$ is defined just below and $C_{R}$ is any lattice path connecting the points $0$ and ${R}$.
Introducing a set of sources ${\zeta}=\{h_1(x), h_2(x), \eta(l) \}$ and integrating out link variables one gets 
\begin{eqnarray}
\Gamma_1({R}) \ = \   \frac{Z_{\Lambda}(\beta,\mu_{\nu};{\zeta})}{Z_{\Lambda}(\beta,\mu_{\nu};0)} \ , \qquad {\zeta}=(h_1(x),0,0) \ ,  
\label{Gamma1}
\end{eqnarray}
where $h_1(x)=1$ for $x=0,{R}$ and $h_1(x)=0$ otherwise and 
\begin{eqnarray}
\Gamma_2({R}) \ = \   \frac{1}{2} \ \frac{Z_{\Lambda}(\beta,\mu_{\nu};\zeta)}{Z_{\Lambda}(\beta,\mu_{\nu};0)} \ + 
\frac{1}{2} \ \frac{Z_{\Lambda}(\beta,\mu_{\nu};\zeta^{\prime})}{Z_{\Lambda}(\beta,\mu_{\nu};0)} \ .   
\label{Gamma2}
\end{eqnarray} 
We have introduced here notations $\zeta=(0,h_2(x),\eta(l))$ and $\zeta^{\prime}=(0,h_2(x),- \eta(l))$, 
where $h_2(x)=1$ for $x=0,{R}$ and $h_2(x)=0$ otherwise;  
$\eta(l)=1$ if {$l=(x;\nu)\in C_R$}, $\eta(l)=-1$ if {$l=(x-e_{\nu};\nu)\in C_R$} and $\eta(l)=0$, otherwise.
The partition function utilized in~(\ref{Gamma1}) and~(\ref{Gamma2}) is given by 
\begin{eqnarray} 
Z_{\Lambda}(\beta,\mu_{\nu};{\zeta})  \ = \ \sum_{q_1=-\infty}^{\infty} \ \sum_{q_2=-\infty}^{\infty} 
e^{-\sum_{\nu=1,2} \ q_{\nu} \mu_{\nu} L_{\nu} - \sum_{l\in C_R} \mu_{\nu}\eta(l)} \nonumber \\ 
\times\sum_{\{r(p) \}=-\infty}^{\infty} \int_0^{\pi} \prod_x \frac{d \alpha(x) }{2} \sin\alpha(x) 
\exp \left [ \beta \sum_{x,\nu}\cos\alpha(x)\cos\alpha(x+e_{\nu}) \right ] \nonumber  \\
\times\prod_x \left ( \cos\alpha(x)  \right )^{h_1(x)}  \ \left ( \sin\alpha(x)  \right )^{h_2(x)} \  \prod_l B_{\eta}(l) \ , 
\label{Zsigma_corr}
\end{eqnarray}
with
\begin{eqnarray}
B_{\eta}(l) \ = \ I_{r(l)+\eta(l)}\left ( \beta \sin\alpha(x)\sin\alpha(x+e_{\nu}) \right ) \ . 
\label{dualB_corr}
\end{eqnarray} 
As it stands, the expression~(\ref{Zsigma_corr}) is much more general and allows to compute correlations of any kind as 
\begin{equation}
\left \langle \ \prod_x \left ( \cos\alpha(x)  \right )^{h_1(x)}  \ \left ( \sin\alpha(x)  \right )^{h_2(x)} \ \prod_l 
\frac{B_{\eta}(l)}{B_0(l)}  \ \right \rangle  \ . 
\label{gen_corr}
\end{equation}

\subsection{Dual Boltzmann weight 1} 

The dual Boltzmann weight can be read off either from~(\ref{Zsigma_mu}) or from~(\ref{Zsigma_corr}). It has the form 
\begin{equation}
e^{-\sum_{\nu=1,2} \ q_{\nu} \mu_{\nu} L_{\nu} } \ \exp \left [ \beta \cos\alpha(x)\cos\alpha(x+e_{\nu}) + 
\log B_0(l)  \right ] \ . 
\label{dual_BW1}
\end{equation}
Clearly, it is strictly positive in the integration domain over $\alpha(x)$ and hence it can be used for the numerical 
simulations. In this case we have 
\begin{equation}
\Gamma_1({R}) \ = \  \langle \ \cos\alpha(0) \cos\alpha({R}) \ \rangle  \ , 
\label{Gamma1_dual1}
\end{equation}
\begin{eqnarray}
\Gamma_2({R}) \ &=& \  \frac{1}{2} \ e^{-\sum_{l\in C_R} \mu_{\nu}\eta(l)} \  \left \langle \ \sin\alpha(0) \sin\alpha({R})  \ 
\prod_{l\in C_{R}} \frac{B_{\eta}(l)}{B_0(l)} \  \right \rangle  \nonumber   \\ 
&+& \frac{1}{2} \ e^{\sum_{l\in C_R} \mu_{\nu}\eta(l)} \  \left \langle \ \sin\alpha(0) \sin\alpha({R})  \ 
\prod_{l\in C_{R}} \frac{B_{- \eta}(l)}{B_0(l)} \  \right \rangle \ . 
\label{Gamma2_dual1}
\end{eqnarray}

\subsection{Dual Boltzmann weight 2} 

A different dual Boltzmann weight can been constructed by performing a complete integration over the remaining original degrees of freedom.

We begin with the formula 
\begin{equation} 
\int_{0}^{\pi} \ d\alpha\ F\left ( \cos\alpha , \sin\alpha \right ) \ = \ 
\sum_{s=\pm 1} \ \int_{0}^{\pi/2} \ d\alpha\ F\left ( s \cos\alpha , \sin\alpha \right )  \ . 
\label{ising_var}
\end{equation}
This introduces {an Ising-like} partition function with fluctuating coupling 
\begin{equation} 
{J}_{\nu}(x) \ = \  \cos\alpha(x)\cos\alpha(x+e_{\nu}) \ , \  \alpha(x) \in [0,\pi/2] \ . 
\label{fluct_coupl}
\end{equation}
Equation~(\ref{Zsigma_corr}) becomes
\begin{eqnarray} 
Z_{\Lambda}(\beta,\mu_{\nu};{\zeta})  =  
\sum_{q_1=-\infty}^{\infty} \ \sum_{q_2=-\infty}^{\infty} \ 
e^{-\sum_{\nu=1,2} \ q_{\nu} \mu_{\nu} L_\nu - \sum_{l\in C_R} \mu_{\nu}\eta(l)} \nonumber \\ 
\times\sum_{\{r(p) \}=-\infty}^{\infty} \int_0^{\pi/2} \prod_x \frac{d \alpha(x) }{2} \sin\alpha(x)
\prod_x \left ( \cos\alpha(x)  \right )^{h_1(x)}  \ \left ( \sin\alpha(x)  \right )^{h_2(x)}\;, \nonumber  \\  
\times\prod_l B_{\eta}(l) \ Z_I(\{{J}_{\nu}(x) \}) \ , 
\label{Zsigma_ising}
\end{eqnarray}
where (taking into account that $h_1(x)=1$ for $x=0,{R}$ and zero otherwise) 
\begin{equation} 
Z_I(\{{J}_{\nu}(x) \}) \ = \ \sum_{\{s(x) \}=\pm 1} \ s(0) s({R}) \ 
\exp\left [ \beta \sum_{x,\nu} \ {J}_{\nu}(x) s(x)s(x+e_{\nu}) \right ] \ .  
\label{Z_ising}
\end{equation}
The dual transformation can be performed in a standard way by introducing the link variables 
$z(l)=s(x)s(x+e_{\nu})$ (this time the global constraints on holonomies can be omitted) to obtain
\begin{equation} 
Z_I(\{{J}_{\nu}(x) \})  =  \sum_{\{z(l) \}=\pm 1}  \prod_{l=1}^{R} z(l) \ 
\exp\left [ \beta \sum_{x,\nu} \ {J}_{\nu}(x) z(l) \right ] 
\prod_p \left [ \sum_{s(p)=0,1} z(p)^{s(p)}  \right ]  , 
\label{Zising_link}
\end{equation}
where $z(p)=\prod_{l\in p}z(l)$. Summation over $z(l)$ leads to the following representation 
on the dual lattice (now $C_{R}$ is a path connecting the points $0$ and ${R}$ and consisting of links 
dual to the original links): 
\begin{eqnarray} 
Z_I(\{{J}_{\nu}(x) \})  =  \sum_{\{ s(x) \}=\pm 1}  \prod_{l\in C_{R}}  
\left [ e^{ \beta {J}_{\nu}(x)} - s(x)s(x+e_{\nu}) e^{ - \beta {J}_{\nu}(x)} \right ] \nonumber  \\ 
\times\prod_{l\notin C_{R}} 
\left [ e^{ \beta {J}_{\nu}(x)} + s(x)s(x+e_{\nu}) e^{ - \beta {J}_{\nu}(x)} \right ] \;.
\label{Zising_dual}
\end{eqnarray}
By inserting~(\ref{Zising_dual}) into~(\ref{Zsigma_ising}), we get the expression
\begin{eqnarray} 
Z_{\Lambda}(\beta,\mu_{\nu};{\zeta})  =  
\sum_{q_1=-\infty}^{\infty} \ \sum_{q_2=-\infty}^{\infty} \ 
e^{-\sum_{\nu=1,2} \ q_{\nu} \mu_{\nu} L_\nu - \sum_{l\in C_R} \mu_{\nu}\eta(l)} \  
\sum_{\{ r(x) \}=-\infty}^{\infty} \sum_{\{ s(x) \}=\pm 1}  \nonumber  \\ 
\label{Zsigma_dual}
\times\int_0^{\pi/2} \prod_p \frac{d \alpha(p) }{2} \ 
\left ( \cos\alpha(p)  \right )^{h_1(p)}  \ \left ( \sin\alpha(p)  \right )^{h_2(p)+1} \ 
\prod_l \ B_{\eta}(l)   \\
\times\prod_{l\in C_{R}}  
\left [ e^{ \beta {J}_{\nu}(x)} - s(x)s(x+e_{\nu}) e^{ - \beta {J}_{\nu}(x)} \right ]  
\prod_{l\notin C_{R}} 
\left [ e^{ \beta {J}_{\nu}(x)} + s(x)s(x+e_{\nu}) e^{ - \beta {J}_{\nu}(x)} \right ] \ . \nonumber 
\end{eqnarray}
Here
\begin{eqnarray}
B_{\eta}(l) \ = \ I_{r(l)+\eta(l)}\left ( \beta \sin\alpha(p)\sin\alpha(p^{\prime}) \right ) \ , 
\label{dualB1}
\end{eqnarray} 
where
\begin{eqnarray} 
r(l) \ = \ 
\begin{cases}
r(x)-r(x+e_{\nu}) + q_{\nu} \ , {\rm if} \ l=(x_1,0;2) \ {\rm or} \ l=(0,x_2;1) \ ,    \\  
r(x)-r(x+e_{\nu}) \ , \ {\rm otherwise} \ . 
\end{cases}
\label{rl_dual}
\end{eqnarray}
The product $\prod_p$ in (\ref{Zsigma_dual}) runs over all plaquettes of the dual lattice (in $2d$ plaquettes are dual to sites and vice-versa). 
Therefore, the definition of $r(l)$ in (\ref{rl}) takes the form of (\ref{rl_dual}) on the dual lattice. The dual plaquettes $p$ and $p^{\prime}$  have the dual link $l$ in common.  
Finally, one can integrate out the $\alpha(p)$ angles. This can be done by Taylor expanding the factor $e^{ \pm \beta {J}_{\nu}(x)}$ 
and by using either the series representation for the Bessel function or the multiplication theorem for the Bessel function 
to decouple the $\sin\alpha(p)$ factors from their argument. The first approach is somewhat simpler and leads to the following result: 
\begin{eqnarray} 
Z_{\Lambda}(\beta,\mu_{\nu};{\zeta})  =  \sum_{q_1,q_2=-\infty}^{\infty} \ e^{-\sum_{\nu=1,2} \ q_{\nu} \mu_{\nu} L_\nu } \  
\sum_{\{ r(x) \}=-\infty}^{\infty} \sum_{\{ s(x) \}=\pm 1}  \ \sum_{\{ m(l),n(l) \}=0}^{\infty} \nonumber  \\ 
\label{Zsigma_dual_fin}
\times\prod_{l}  \frac{\beta^{n(l)}}{n(l)!}  \ 
\frac{ \left ( \frac{\beta}{2} \right )^{2m(l)+|r(l)+\eta(l)|}e^{-\mu_{\nu}\eta(l)}}  
{m(l)!(m(l)+|r(l)+\eta(l)|)!} \ \prod_p B\left ( \frac{a(p)+1}{2}, \frac{b(p)+1}{2}  \right ) \\
\times\prod_{l\in C_{R}}  
\left [ 1 - s(x)s(x+e_{\nu}) (-1)^{n(l)} \right ]  
\prod_{l\notin C_{R}} 
\left [ 1 + s(x)s(x+e_{\nu}) (-1)^{n(l)} \right ] \ , \nonumber 
\end{eqnarray}
where $B(x,y)$ is the beta-function and $a(p)$ and $b(p)$ are given by   
\begin{eqnarray}
a(p) &=& \sum_{l\in p} n(l) + h_1(p)   \ , \nonumber   \\ 
b(p) &=& {1+}\sum_{l\in p} \left ( 2m(l) + |{ r(l)}+\eta(l)| \right ) + h_2(p) \ . 
\label{const_ab}
\end{eqnarray}

The partition function can be obtained from the last expression if we put $h_1(p)=h_2(p)=\eta(l)=0$ and 
extend the second product in the last line to all links of the lattice 
\begin{eqnarray} 
Z_{\Lambda}(\beta,\mu_{\nu})  =  \sum_{q_1,q_2=-\infty}^{\infty} \ e^{-\sum_{\nu=1,2} \ q_{\nu} \mu_{\nu} L_\nu} \  
\sum_{\{ r(x) \}=-\infty}^{\infty} \sum_{\{ s(x) \}=\pm 1}  \ \sum_{\{ m(l),n(l) \}=0}^{\infty} \nonumber  \\ 
\label{Zsigma_dual_finpf}
\times\prod_{l}  \frac{\beta^{n(l)}}{n(l)!}  \ 
\frac{ \left ( \frac{\beta}{2} \right )^{2m(l)+|r(l)|}}  
{m(l)!(m(l)+|r(l)|)!} \ \prod_p B\left ( \frac{a(p)+1}{2}, \frac{b(p)+1}{2}  \right ) \\
\times\prod_l  \left [ 1 + s(x)s(x+e_{\nu}) (-1)^{n(l)} \right ] \ . \nonumber 
\end{eqnarray}
Here, $r(l)$ is given in (\ref{rl_dual}) and  
\begin{eqnarray}
a(p) &=& \sum_{l\in p} n(l)   \ , \nonumber   \\ 
b(p) &=&{1+} \sum_{l\in p} \left ( 2m(l) + |r(l)| \right )  \ . 
\label{apbp_def}
\end{eqnarray}
The Boltzmann weights of all three representations~(\ref{Zsigma_mu}), (\ref{Zsigma_dual}) and~(\ref{Zsigma_dual_finpf})
are positive and all interactions between dual variables are local. 
Moreover, the dual weight of (\ref{Zsigma_mu}) is free of constraints. 
It means, in particular, that with the help of (\ref{Zsigma_mu}) one can compute not only 
local quantities (those which can be {represented} as derivatives of the partition function), but also long-distance quantities 
like the two-point correlation function, because it can be {represented} as an ordinary expectation value.

\section{Simulation details} 
\label{simdetails}

We have derived two different real Boltzmann weights, (\ref{Zsigma_mu}) and (\ref{Zsigma_dual_finpf}), for the same model. However, being real is not
the only condition that an action must satisfy to be handy during numerical simulations. In conjunction with an adequate simulation algorithm, the action should
also avoid slowing down. Therefore, and in order to elucidate the efficiency of the two weights, we have tested both (\ref{Zsigma_mu}) and (\ref{Zsigma_dual_finpf}).
The results of physical magnitudes will be presented in the next section but we anticipate that both weights exhibit similar acceptances and simulation efficiencies.
{The only difference that is worth stressing regards the computation of correlation functions: it is much more problematical with (\ref{Zsigma_dual_finpf}) than with (\ref{Zsigma_mu})
because, as is evident from expressions (\ref{Zsigma_dual_fin}) and (\ref{Zsigma_dual_finpf}), the correlation function must be determined as the ratio of both quantities and
such ratios are usually so noisy that it is computationally very expensive to prevent the error bars from growing excessively. For all of that,
once we verified that the performances of the two weights are equivalent in every respect, we have employed only (\ref{Zsigma_mu})
in the battery of Monte Carlo simulations aimed at extracting physical properties of the model.}

We have simulated the weight (\ref{Zsigma_mu}) on square lattices $\Lambda\in{\mathbb{Z}}^2$ of lateral extent\footnote{{Whenever we write $L$, we will mean that $L_1=L_2\equiv L$.}}
$L_1=L_2\equiv L$ with periodic boundary conditions.  The variables are $\alpha(x)\in[0,\pi]$, $r(p)\in\mathbb{Z}$ and $q_1,q_2\in\mathbb{Z}$ where $x$ indicates sites and $p$ plaquettes.
A single {Monte Carlo sweep} consists in updating every variable $\alpha(x)$, $r(p)$ and $q_1,q_2$ {with the Metropolis algorithm~\cite{metropolis} once.}

{The refreshing of the angle variables $\alpha(x)$ was done by proposing} a brand new value of $\cos\alpha(x)$ { with equal probability from the interval $[-1,+1]$,} and applying the usual Metropolis test on it.
Plaquette variables $r(p)$ and global variables $q_1,q_2$ were updated after randomly choosing a new value that differs from the old one by at most $\pm\Delta$ units (we took { $\Delta=3$} in $r(p)$ and $q_1$, $q_2$).

\begin{figure}[!t]
\centering
\includegraphics*[width=120mm]{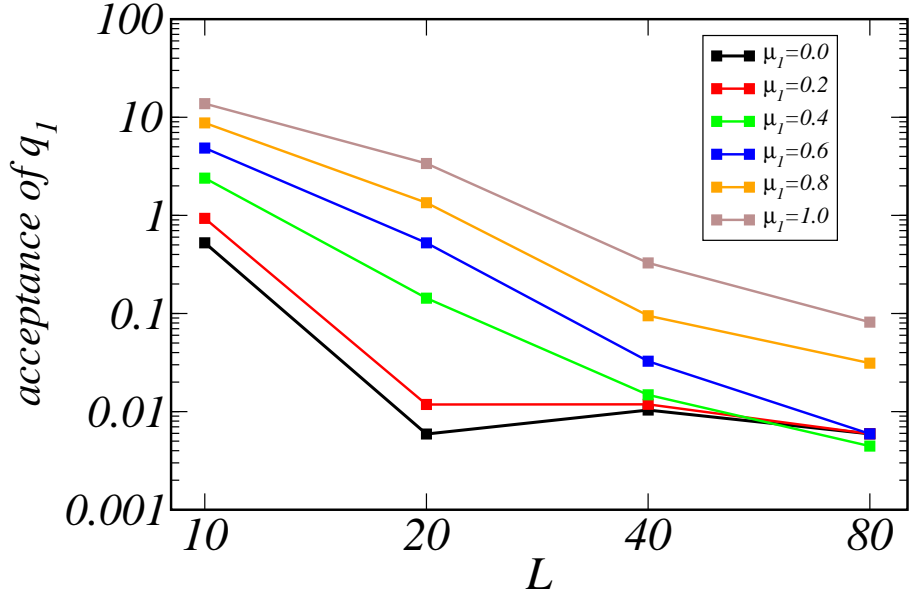}
\caption{(Color online) Acceptance rates of variable $q_1$ for $\beta=1.2$ at $\mu_2=0$ and for the indicated values of $\mu_1$ and of the lattice size. The lines are guides to the eye.}
    \label{fig_accept.pdf}
\end{figure}

{The acceptance rate of a dynamical variable is defined as the percentage of these variables that are changed on average during the Monte Carlo sweeps. These rates}
were generally quite low, particularly for $q_\mu$. After setting $\mu_2=0$,
we show in Fig.~\ref{fig_accept.pdf} the acceptances for $q_1$ as a function of $\mu_1$ and of the lattice size $L$ for $\beta=1.2$. {They exhibit a downward trend as $L$ or $1/\mu_1$ grow, possibly
following a power law behaviour. Such a behaviour will provoke a sudden growth of the error bars for any observable as $L$ and $1/\mu_1$ increase beyond certain values.}

{The computational times needed to gather $2\cdot10^7$ measurements on a $20^2$ lattice is about 30 hours on a node with four processors of the type AMD Opteron(tm) Processor 6376.}

Next we introduce the physical observables measured in this work. In order to enhance decorrelation, single measurements were evaluated on configurations separated by
{10~Monte Carlo sweeps}. Error bars were assessed by the Jackknife method applied on blocks of data for further reducing any correlation among raw
  data. We typically considered ten levels of blocking, the number of blocks ranging from a minimum of ten to a maximum of a few thousands.

\begin{figure}[!t]
  \centering
  \includegraphics*[width=120mm]{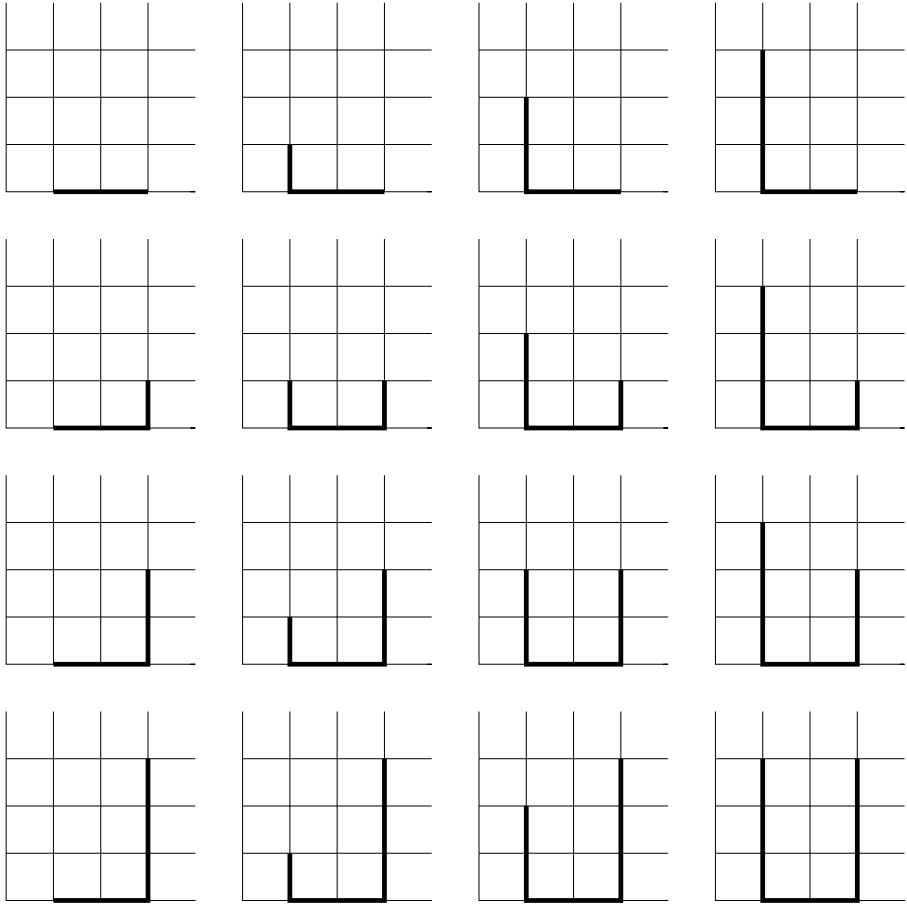}
  \caption{Paths of a wall-wall correlation function at distance~2 on a $4\times4$ lattice. The walls are erected at sites $(2,0)$ and $(4,0)$.}
      \label{wallwall_bis.pdf}
\end{figure}

The measured magnitudes are:

\begin{enumerate}[label=(\roman*)]

\item the correlation length which was measured by analysing the wall-wall correlation function. This function was constructed according to the definition (\ref{Gamma2_dual1}).
  Since the path followed to join the walls is irrelevant (on average), we chose it as the simplest one guaranteeing computing efficiency: along the $x_1$-axis. For clarity, in Fig.~\ref{wallwall_bis.pdf} the~16
  paths composing a wall-wall correlation at distance~2 from site $(2,0)$ to site $(4,0)$ on a $4\times4$ lattice are shown. The total wall-wall correlation function at distance~2
  consists in summing the contributions from all the above paths and averaging the result over all possible ways the walls can be placed at distance~2: erecting the walls at sites $(1,0)$ and $(3,0)$, at $(2,0)$ and $(4,0)$
  ---this is the one shown in Fig.~\ref{wallwall_bis.pdf}---, at $(3,0)$ and $(1,0)$,
  and at $(4,0)$ and $(2,0)$, having used periodicity on the last two. In general, that average contains $L$ terms on a $L\times{L}$ lattice;

\item the energy density $E$ which is defined as
  \begin{equation}
    E=\frac{1}{2L_1L_2}\frac{\partial \ln Z}{\partial \beta}\;,
    \label{Edef}
  \end{equation}
and given by
      \begin{equation}
        E = \frac{1}{2L_1L_2} \ \sum_{x,\nu}
        \langle \ \cos\alpha(x)\cos\alpha(x+e_\nu)
        + \frac{I_{r(l)+1}(\beta \gamma(l))+I_{r(l)-1}(\beta\gamma(l))}{2I_{r(l)}(\beta\gamma(l))}\, \gamma(l)\ \rangle \;,
        \label{energy}
      \end{equation}
 where $r(l)$ is defined in (\ref{rl}) and $\gamma(l)$ is $\gamma(l)\equiv\sin\alpha(x)\sin\alpha(x+e_\nu)$;

\item the particle density. Since the number $N$ of particles is
  \begin{equation}
    N=\frac{\partial \ln Z}{\partial \mu_1}=-\langle L_1 q_1\rangle\;,
  \end{equation}
  we deduce that the particle density is 
  \begin{equation}
    n\equiv - \frac{1}{L_2}\,\langle q_1\rangle\;,
    \label{densityn1}
  \end{equation}
  or, equivalently,
    \begin{equation}
    n\equiv - \frac{1}{L_1}\,\langle q_2\rangle\;.
    \label{densityn2}
    \end{equation}
Both observables were measured for testing purposes as both~(\ref{densityn1}) and~(\ref{densityn2}) should provide the same result. 
The test was successfully passed.

\end{enumerate}

\section{Results}
\label{results}

{The Monte Carlo simulations have been carried out on square lattices $L_1=L_2\equiv L$ with $L=20$, $30$, and~40. $\mu_2$ was always set to zero and we will call $\mu\equiv\mu_1$.
  Having not too low acceptances is an indispensable condition to regard the results in every simulation run as valid. Therefore, since as shown in Fig.~\ref{fig_accept.pdf}, our algorithm slowed down
  on large lattice sizes, we were compelled to employ $L\leq40$. Fig.~\ref{fig_accept.pdf} indeed exhibits the poor variability of $q_1$.} We typically collected some $10^7$ measurements for each run.


Our strategy to construct the dual Boltzmann weight (\ref{Zsigma_mu}) allows to calculate every observable, including correlation functions.
Such functions permit to extract the mass gap. This gap { should coincide} (i) with the value of the chemical potential $\mu$ at which
non-zero particle numbers start off and (ii) with the value of $\mu$ at which the energy density detaches from its value at zero particle
density. Therefore our procedure enables us to cross-check the results for the mass gap.

\begin{figure}[!t]
  \centering
  \includegraphics*[width=120mm]{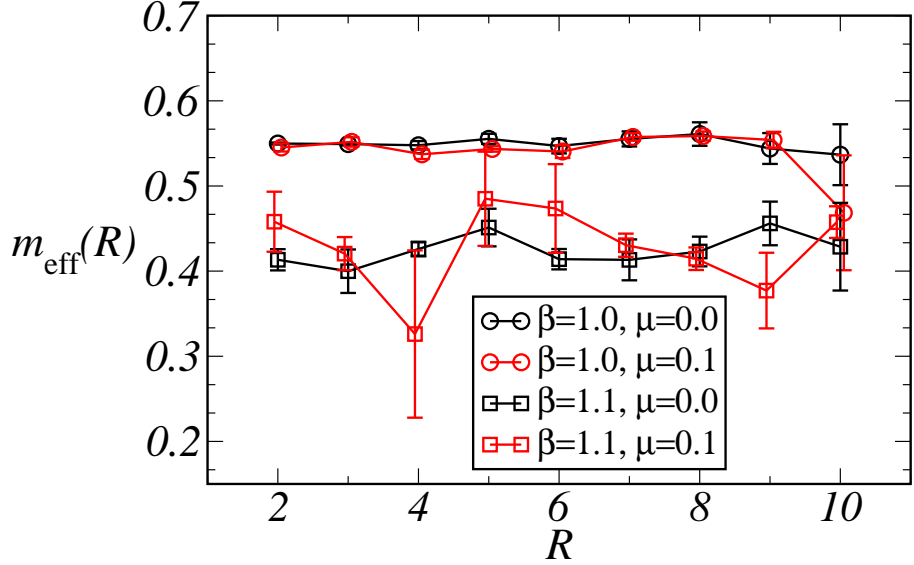}
  \caption{(Color online) Effective mass extracted from the correlator at different distances and values of $\mu$ for $\beta=1.0$ and $\beta=1.1$ on a $20\times20$ lattice. Lines are
      guides to the eye. The independence on $\mu$ is evident, as well as the presence of plateaux. All measurements have been done at integer values of {$R$} (but some of them appear slightly shifted for readability).}
      \label{fig_mass.pdf}
\end{figure}

Assuming the following theoretical form for the wall-wall correlation function,
\begin{equation}
\Gamma^{\rm (th)}_{\rm w}(R) = A \bigl[ e^{-Rm} + e^{-(L-R)m}\bigr]\;,
\end{equation}
we extracted the {\em effective mass} {$m_{\rm eff}(R)$} as the value of the parameter $m$ at which the equality
\begin{equation}
-\log\left[\frac{\Gamma^{\rm (MC)}_{\rm w}(R+1)}{\Gamma^{\rm (MC)}_{\rm w}(R)}\right] =
-\log\left[\frac{\Gamma^{\rm (th)}_{\rm w}(R+1)}{\Gamma^{\rm (th)}_{\rm w}(R)}\right]
\end{equation}
holds, where $\Gamma^{\rm (MC)}_{\rm w}(R)$ stands for the Monte Carlo determination of the wall-wall correlation. 
Here $m_{\rm eff}(R)$ is given in units of the inverse lattice spacing. We expect that $m_{\rm eff}(R)$
exhibits a plateau for large enough $R$. The height of that plateau is the value of the mass gap resulting from our simulations. We call $m_{\rm MC}$ this height.
In Fig.~\ref{fig_mass.pdf} we summarize our findings for two values of $\beta$ and $\mu$. This plot follows after measuring the correlation function on $2\cdot10^7$ configurations
obtained in a $20\times20$ lattice. Similar results have been derived for other $\beta$ and $\mu$ (not shown). In all cases the expected independence on the chemical potential is apparent.

\begin{figure}[!t]
  \centering
  \includegraphics*[width=120mm]{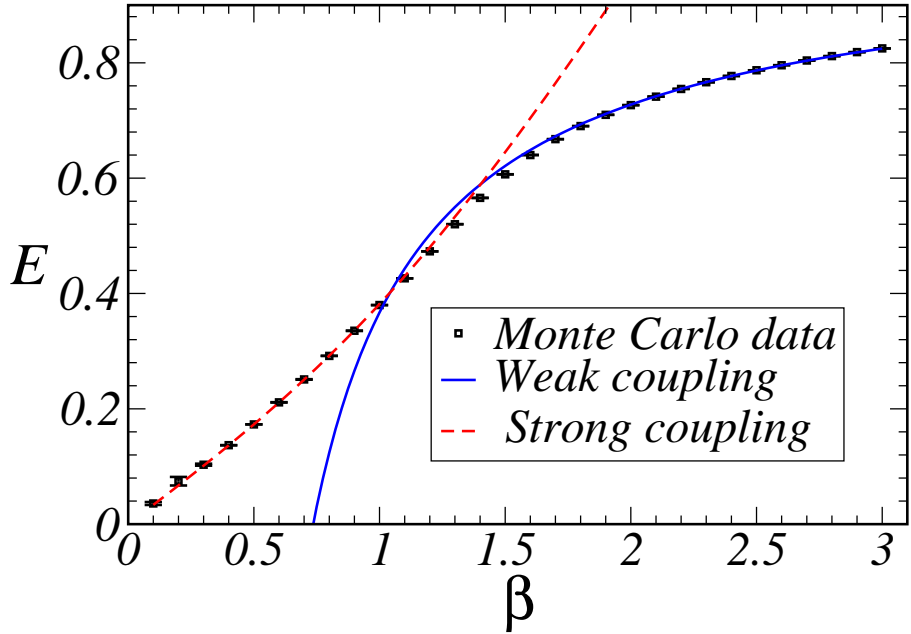}
  \caption{(Color online) Energy as a function of $\beta$ for $\mu=0$ on a $20\times20$ lattice. Monte Carlo data are represented by symbols $\Box$.
    The result is in perfect agreement with the weak coupling and strong coupling expansions, represented respectively by continuous and dashed lines.}
  \label{fig_Ener_vs_b.pdf}
\end{figure}

In Fig.~\ref{fig_Ener_vs_b.pdf} the energy density~(\ref{energy}) for zero particle density is displayed as a function of {$\beta$}. Data in this figure are obtained again on a $L=20$ lattice.
They exhibit a satisfactory agreement with the weak coupling expansion\footnote{The coefficient of $1/\beta^4$ includes the slight correction found in \cite{pepe}.}\cite{Alles97}
\begin{equation}
  E=1-\frac{1}{2\beta}-\frac{1}{16\beta^2}-\frac{0.03851}{\beta^3}-\frac{0.03189}{\beta^4}+\cdots\;,
  \label{weaktomeu}
\end{equation}
and with the strong coupling expansion~\cite{berg81}
\begin{equation}
  E=y+2 y^3 + \frac{12}{5} y^5+\cdots\;,\qquad y\equiv\frac{1}{\tanh \beta}-\frac{1}{\beta}\;.
  \label{otrostrong}
\end{equation}
These agreements constitute further positive tests of our procedure.

\begin{figure}[!t]
  \centering
  \includegraphics*[width=120mm]{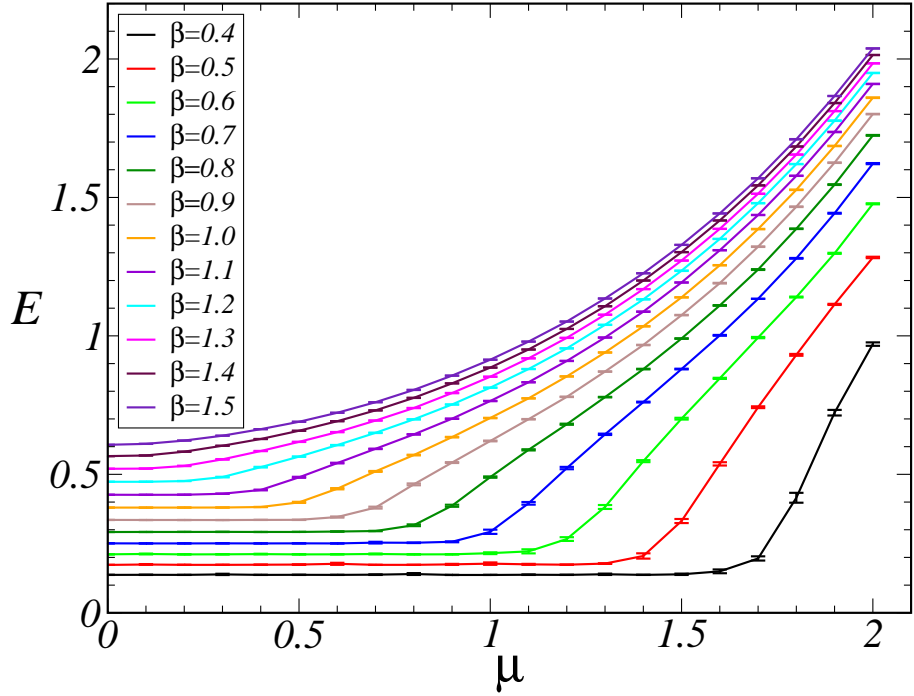}
  \caption{(Color online) Energy as a function of $\mu$ and $\beta$ on a $20\times20$ lattice. Lines are guides to the eye. The non-trivial dependence on $\mu$ starts only at $\mu=m_{\rm MC}$.}
  \label{fig_Ener_vs_mb.pdf}
\end{figure}

Figure~\ref{fig_Ener_vs_mb.pdf} shows the energy density $E$ as a function of $\beta$ and $\mu$ on a $20\times20$ lattice. As expected, the value of $\mu$ at which $E$ detaches from
its value at $\mu=0$ is $\mu=m_{\rm MC}$. Even though we have used rather small lattice sizes, the coincidence between thresholds and effective mass $m_{\rm MC}$
is manifest because the energy density operator has negligible size effects.

\begin{figure}[!t]
  \centering
  \includegraphics*[width=120mm]{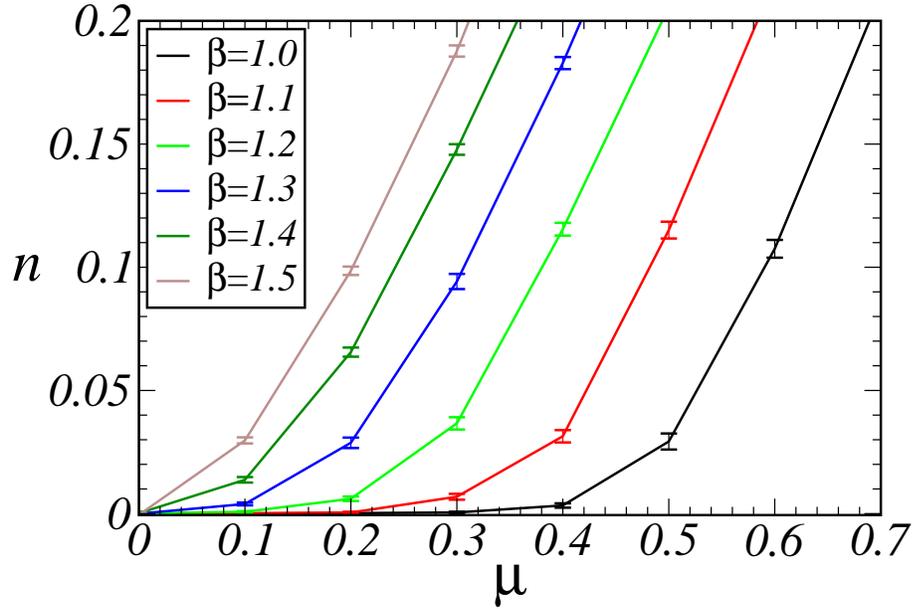}
  \caption{(Color online) Particle density as a function of $\mu$ and $\beta$ on a $20\times20$ lattice. Lines are guides to the eye.
    The threshold value for $\mu$ corresponds to the mass gap in units of the inverse lattice spacing and coincides with the thresholds shown in Fig.~\ref{fig_Ener_vs_mb.pdf}.}
  \label{fig_dens_vs_mb.pdf}
\end{figure}

\begin{figure}[!t]
  \centering
  \includegraphics*[width=120mm]{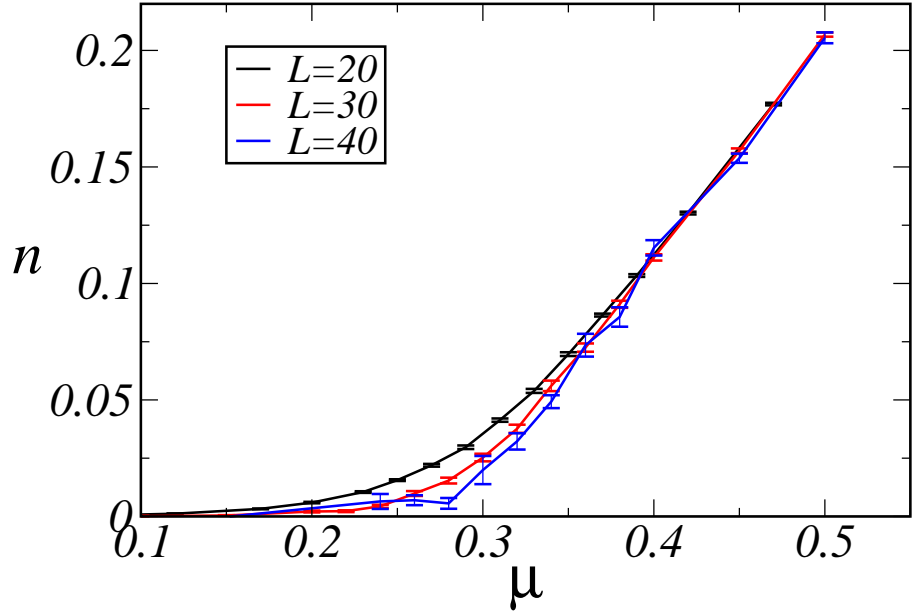}
  \caption{(Color online) Particle density as a function of $\mu$ at $\beta=1.2$ for $L=20,\,30,\,40$. Lines are guides to the eye.
    The threshold is evidently size dependent.}
  \label{fig_dens_vs_m_b1.2.pdf}
\end{figure}

Figure~\ref{fig_dens_vs_mb.pdf} shows the dependence of the particle density on $\mu$ for several $\beta$. Each point is the average of $10^6$ measurements.
In principle also the thresholds in this plot should coincide with the plateaux in Fig.~\ref{fig_mass.pdf}. This coincidence is not so manifest at $L=20$ because, contrary to the energy density, $n$ has
strong size effects~\cite{niedermayer}. We have verified this assertion by repeating the study for $L=30$ and $L=40$ at $\beta=1.2$. Figure~\ref{fig_dens_vs_m_b1.2.pdf}
  blatantly exhibits that size dependence. While the threshold at $L=20$ is at about $\mu\approx0.2$, at $L=40$ it is shifted to about $\mu\approx0.3$.
  The point along the line for $\beta=1.2$ in Figure~\ref{fig_Ener_vs_mb.pdf} at which data detach from the value of the energy at zero density agrees reasonably well
  with the threshold in Figure~\ref{fig_dens_vs_m_b1.2.pdf} for $L=40$.

  Figure~\ref{fig_dens_vs_m_b1.2.pdf} also displays the consequences of the above-mentioned slowing down. Indeed, although the three lines have been
  obtained with the same statistics, namely $2\cdot10^7$ measurements, all points on the line for $L=40$ and some of the points along the line for $L=30$
  present larger error bars. They are due to long correlations among data. Such errors can be reduced only at the cost of increasing exorbitantly the statistics.

\section{Discussion}
\label{discus}

We have derived two different positive Boltzmann weights, (\ref{Zsigma_mu}) and (\ref{Zsigma_dual_finpf}), for simulating the $2d$ $O(3)$ non-linear sigma model 
at non-zero chemical potential. {These weights permit the evaluation of any observable average.}
The performances of both weights during Monte Carlo simulations are similar. For example in a wide region of $\beta$-$\mu$ the simulations run correctly, but low acceptances
in the dynamical variables occur for both actions under certain values of $\beta$, $\mu$ and lattice sizes. {This fact prevents the use of large lattices to avoid finite size
effects, mainly at low, but non-zero, chemical potentials. Tempered Monte
Carlo did not manage to speed up} the simulation at those lattice sizes. Clearly all that results in strong slowing down effects on the simulation.

\begin{figure}[!t]
  \centering
  \includegraphics*[width=120mm]{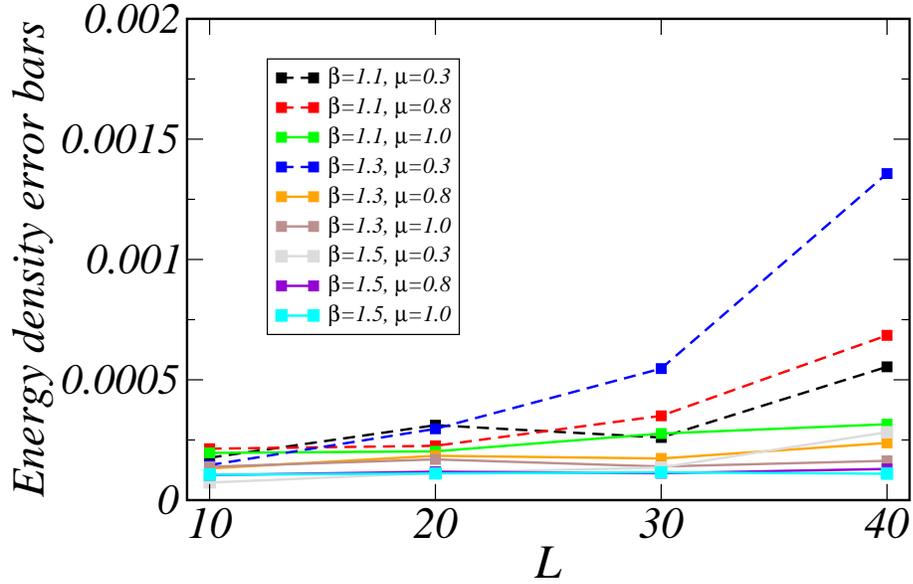}
  \caption{(Color online) {Error bars for the energy density from a sample of $10^7$ data as a function of $L$. Lines are guides to the eye.
    Data exhibiting a steep growth of the error bars have been represented with dashed lines.}}
  \label{errors_E.pdf}
\end{figure}

\begin{figure}[!t]
  \centering
  \includegraphics*[width=120mm]{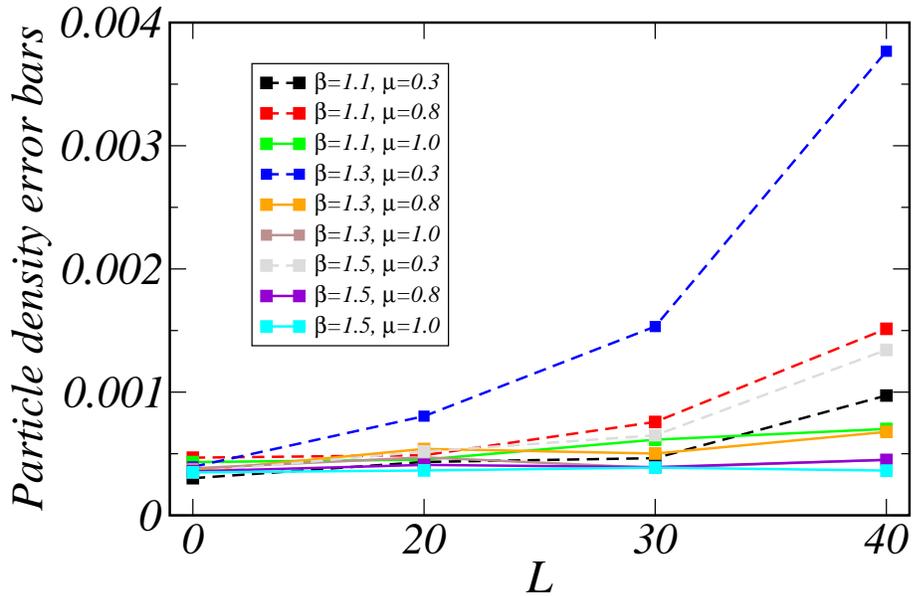}
  \caption{(Color online) {Error bars for the particle density from a sample of $10^7$ data as a function of $L$. Lines are guides to the eye.
    Data exhibiting a marked growth of the error bars have been represented with dashed lines.}}
  \label{errors_ndens.pdf}
\end{figure}

In turn, the above slowing down makes the values of the error bars to become large. This phenomenon is shown in Fig.~\ref{errors_E.pdf} and Fig.~\ref{errors_ndens.pdf} for
the error bars of the energy and of the particle density. Such a growth raises the suspect that the sign problem may have not been completely beaten, even though the actions
(\ref{Zsigma_mu}) and (\ref{Zsigma_dual_finpf}) derived in the paper are manifestly real. However, the fact that this slowing down looks more intense for low chemical
potentials than for large ones (see Fig.~\ref{fig_accept.pdf}) seems to indicate that its origin has nothing to do with the old sign problem (which worsens as $\mu$ increases)
and is simply a spurious consequence of our procedure.

A detailed study of the type of functional growth of the error bars is hindered by the onset of the above-described slowing down at lattice sizes larger than~40.

The above difficulties are absent on not too large lattice sizes. Thus, it is in those restrained sizes that we have measured all physically meaningful quantities, the sizes ranging from
10 to 40. Specifically, we have extracted the correlation functions and from them the mass gap. These calculations have been confronted with the determination of the
mass gap from the behaviour of the particle and energy densities as functions of $\mu$. The comparison has been successful.
Also the energy density matches the weak and strong coupling expansions (where they should) for $\mu=0$.

It remains to investigate and solve the above slowing down and to perform a more accurate study of the model on larger lattices (which would enable us
to address the question of an hypothetical BKT transition in $O(N)$ models at finite temperature and non-vanishing chemical potential). 
Appealing is also the possibility to extend the procedure described here to the $2d$ $O(3)$ non-linear sigma model with a topological 
$\theta$-term. In this case the sign problem is even more severe. In the past we {derived} a positive weight for the model at non-zero $\theta$ 
by mapping it to a certain dual version of the $SU(2)$ principal chiral model~\cite{dnepr_conf,lattice_13}. However, the worm algorithm 
applied for this dual model performed rather poorly during the Monte Carlo tests~\cite{lattice_13}. 
The approach described in this paper can be generalized to the model which includes the $\theta$-term. Whether or not one can construct 
the positive Boltzmann weight in this case or, at least, significantly reduce the sign problem remains to be verified.

\vspace{0.5cm}

{\bf \large Acknowledgements}

\vspace{0.2cm}

{ We gratefully acknowledge several illuminating discussions with Michele Caselle, Volodymyr Chelnokov and Marco Rossi.
{O.B. and B.A. also thank the Department of Physics of the University of Calabria for the warm hospitality during several stays.} O.B.
also thanks INFN for financial support. Numerical simulations have been run on the ReCaS Data Center of INFN-Cosenza.}

\end{document}